\documentclass[aps,prb,superscriptaddress,twocolumn,showpacs,10pt]{revtex4-1}
\pdfoutput=1
\usepackage{amsmath,amssymb,mathrsfs,bbold}
\usepackage{graphicx,color,ulem}
\usepackage{bbm}

\newcommand{\dunderline}[1]{\underline{\underline{#1}}}

\allowdisplaybreaks[4]

\begin{document}
%%%%%%%%%%%%%%%%%%%%%%%%%%%%%%%%%%%%%%%%%%%%%%
\title{{\begin{center}The generic fixed point model for 
pseudo-spin-$\frac{1}{2}$ quantum dots in nonequilibrium:
\\
Spin-valve systems with compensating spin polarizations\end{center}}}
%%%%%%%%%%%%%%%%%%%%%%%%%%%%%%%%%%%%%%%%%%%%%%
\author{Stefan G\"ottel}
\author{Frank Reininghaus}
\author{Herbert Schoeller}
\affiliation{Institute for Theory of Statistical Physics, RWTH Aachen,
52056 Aachen, Germany}
\affiliation{JARA-Fundamentals of Future Information Technology}

\date{\today}
\pagestyle{plain}

\begin{abstract}
We study a pseudo-spin-$\frac{1}{2}$ quantum dot in the cotunneling regime close
to the particle-hole symmetric point. For a generic tunneling matrix 
we find a generic fixed point with interesting nonequilibrium properties, 
characterized by effective reservoirs with compensating spin orientation vectors 
weighted by the polarizations and the tunneling rates. At large bias voltage we study
the magnetic field dependence of the dot magnetization and the current. 
The fixed point can be clearly identified by analyzing the magnetization of the dot.
We characterize in detail the universal properties for the case of two reservoirs.
\end{abstract}
%05.60.Gg   Quantum transport
%71.10.-w   Theories and models of many-electron systems
%72.10.Bg   General formulation of transport theories
%73.23.-b   Electronic transport in mesoscopic systems
%73.63.Kv   Quantum dots 
%\keywords{}

\maketitle
Nonequilibrium properties of strongly interacting quantum dots have gained an enormous interest
in the last decades. Quantum dots are experimentally controllable systems 
useful for a variety of applications in nanoelectronics, spintronics and 
quantum information processing \cite{review_qi}. They are of fundamental interest
in the field of open quantum systems in nonequilibrium with interesting quantum  
many-body properties and coherent phenomena at low temperatures \cite{andergassen_etal}. 
Of particular interest are spin-dependent phenomena where
the quantum dot is tuned to the Coulomb blockade regime. In the case of a singly-occupied 
dot the spin can fluctuate between two values leading to a realization of the 
isotropic spin-$\frac{1}{2}$ antiferromagnetic Kondo model. A hallmark was the 
prediction and observation of universal conductance for this model
\cite{kondo_theo,kondo_exp}. The equilibrium properties of the Kondo model have been studied extensively 
\cite{costi_etal_94,glazman_pustilnik_05} and, most recently, by using renormalization 
group (RG) methods in nonequilibrium, also the properties at finite bias voltage and the
time dynamics have been analyzed in weak
\cite{rosch_etal,kehrein_etal,schoeller_epj09,hs_reininghaus_prb09,pletyukhov_etal_prl10} 
and strong coupling \cite{frg_noneq,strong_coupling_RTRG,smirnov_grifoni_03} and compared to experiments 
\cite{strong_coupling_exp}. 

The isotropic Kondo model with unpolarized leads is only a special case out of the whole class
of quantum dot models where a single particle on the dot can fluctuate between two different quantum numbers
(which we call a pseudo-spin-$\frac{1}{2}$ quantum dot in the following).
Besides the case of ferromagnetic leads with arbitrary spin orientations 
the two quantum numbers can also label two different orbitals 
or can arise from a mixture of spin and orbital degrees of freedom in the presence of spin-orbit 
interaction in the leads or on the dot, leading to non-spin-conserving tunneling matrices. 
In equilibrium (or the linear response regime), it has been 
found for several cases that exchange fields are generated but if those 
are canceled by external ones the universality properties of the Kondo model are re-established. 
This has been confirmed by numerical renormalization group (NRG) calculations for 
ferromagnetic leads with parallel or antiparallel orientations
\cite{koenig_nrg} and for quantum dots with orbital degrees of freedom or Aharonov-Bohm geometries
\cite{boese_etal}. In Ref.~\onlinecite{kashcheyevs_etal_prb07} a mapping between these different models 
and an analytical understanding in terms of the anisotropic Kondo model has been established.
Concerning nonequilibrium transport previous studies have focused on exchange fields
generated by ferromagnetic leads \cite{koenig_etal}, spin-orbit interaction 
\cite{paaske_etal_prb10} or orbital fluctuations \cite{boese_etal}. 
A systematic nonequilibrium RG study of a pseudo-spin-$\frac{1}{2}$ quantum dot with spin-orbit interaction in the
cotunneling regime has been performed in Ref.~\onlinecite{pletyukhov_schuricht_prb11}, where a 
Dzyaloshinskii-Moriya (DM) interaction together with exchange fields proportional to the bias 
voltage have been identified. For special orientations of the DM-vectors interesting asymmetries 
in resonant transport where reported when a magnetic field of the order of the bias voltage is applied.

\begin{figure}[t]
  \centering
    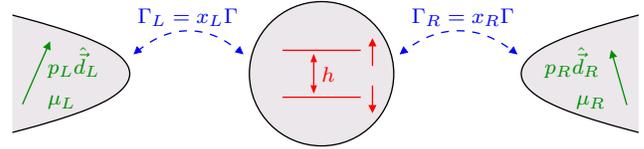
  \caption{(Color online)
  Sketch of the effective model of two ferromagnetic leads $\alpha=L,R$ coupled to  
  a pseudo-spin-$\frac{1}{2}$ quantum dot via spin-conserving tunneling rates $\Gamma_{L,R}=x_{L,R}\Gamma$. 
  $\mu_{L,R}=\pm V/2$ denote 
  the chemical potentials of the leads with spin axis $\hat{\vec{d}}_{L,R}$ and spin polarization 
  $p_{L,R}$. $h$ denotes the Zeeman splitting of the dot levels including exchange fields.}
  \label{fig:model}
\end{figure}

All previous references treated special cases of pseudo-spin-$\frac{1}{2}$ quantum dots without aiming at
finding generic features common to all these systems, irrespective of the complexity of
the geometry, the special interactions and the polarizations of the reservoirs. 
The purpose of this letter is to establish such features especially in the nonequilibrium
regime. Thereby, we will first use a mapping to a pseudo-spin-$\frac{1}{2}$ quantum dot coupled
to effective ferromagnetic leads as depicted in Fig.~\ref{fig:model}, similiar to 
Refs.~\cite{koenig_etal,kashcheyevs_etal_prb07}. Based on this model, we will show that in
the Coulomb blockade regime close to the particle-hole symmetric point a fixed point model 
can be identified where the average of the unit vectors of the spin orientations 
$\hat{\vec{d}}_\alpha$ weighted by the polarizations $p_\alpha$ and the tunneling rates 
$\Gamma_\alpha$ compensate each other ($\alpha$ is the reservoir index)
\begin{align}
\label{eq:univ_prop}
\vec{d}=\sum_\alpha \vec{d}_\alpha = 0 \quad,\quad
\vec{d}_\alpha = x_\alpha p_\alpha \hat{\vec{d}}_\alpha  \quad,\quad
x_\alpha=\frac{\Gamma_\alpha}{\Gamma}\quad,
\end{align}
with $\Gamma=\sum_\alpha\Gamma_\alpha$. This explains why the Kondo effect appears generically 
in the equilibrium case where all reservoirs can be taken together and 
\eqref{eq:univ_prop} leads to a vanishing spin polarization, in agreement with
Rfs.~\onlinecite{koenig_nrg,boese_etal,kashcheyevs_etal_prb07}. However, what
has been overlooked so far is that the fixed point model is generically {\it not} the one of the Kondo model
with one unpolarized lead but rather a spin-$\frac{1}{2}$ coupled to several leads with {\it different} 
spin vectors $\vec{d}_\alpha$. This is particularly important for the nonequilibrium case where
the reservoirs cannot be taken together. Thus, an interesting fixed point emerges which, in the 
equilibrium case, leads to the usual Kondo physics, whereas, in the nonequilibrium regime, 
shows essentially {\it different} universal behavior compared to the Kondo model. We will characterize the
universal features by calculating the magnetic field dependence of the dot magnetization 
and the charge current at zero temperature and large chemical potentials $\mu_\alpha$ compared to the Kondo
temperature $T_K$ at and away from the fixed point. As a smoking gun to detect the fixed point we
find that the dot magnetization $\vec{M}=\langle\vec{S}\rangle$ 
is minimal for all magnetic fields lying on a sphere defined by
\begin{align}
\label{eq:smoking_gun}
|\vec{h}-\vec{\mu}|=|\vec{\mu}|\quad,\quad
\vec{\mu}=\sum_\alpha (\mu_\alpha-\bar{\mu})\vec{d}_\alpha\quad,
\end{align}
where $\bar{\mu}=\sum_\alpha x_\alpha\mu_\alpha$. We note that $\vec{h}$ denotes the total
magnetic field including exchange fields. We choose units $\hbar=e=1$.

{\it Effective model.}
We start from a generalized Anderson impurity model, where the dot Hamiltonian is given by
$H=\sum_{\sigma}\epsilon_{\sigma}n_\sigma + U n_\uparrow n_\downarrow$, where 
$\epsilon_\sigma=\epsilon + \sigma h/2$ are the single-particle energies and 
$U$ denotes a strong Coulomb repulsion. The dot
is coupled to the reservoirs by a generic tunneling matrix 
$(\dunderline{t}{}_\alpha)_{\nu\sigma}=t^\alpha_{\nu\sigma}$, where $\nu$ is a channel index
labelling the reservoir bands with possibly different density of states (d.o.s.) $\rho_{\alpha\nu}$ (in
dimensionless units). The key observation is that the reservoirs enter only via the retarded
self-energy, which is fully characterized by the hybridization matrix
$\dunderline{\Gamma}{}_{\alpha}=2\pi\,\dunderline{t}{}_\alpha^\dagger
\dunderline{\rho}{}_\alpha\dunderline{t}{}_\alpha$, with
$(\dunderline{\rho}{}_\alpha)_{\nu\nu'}=\rho_{\alpha\nu}\delta_{\nu\nu'}$. 
This means that all models with the same matrix $\dunderline{\Gamma}{}_\alpha$ give the same 
result for the dot density matrix and the charge current. 
Once $\dunderline{\Gamma}{}_\alpha$ is known, we can write it in various forms to obtain 
effective models. $\dunderline{\Gamma}{}_\alpha$ is a positive semidefinite Hermitian $2\times 2$-matrix, 
i.e., it can be diagonalized by a unitary $2\times 2$-matrix $\dunderline{U}{}_\alpha$ such that
$\dunderline{\Gamma}{}_{\alpha}=
\dunderline{U}{}_\alpha^\dagger\dunderline{\tilde{\Gamma}}{}_\alpha\dunderline{U}{}_\alpha$
with the diagonal matrix 
$(\dunderline{\tilde{\Gamma}}{}_{\alpha})_{\sigma\sigma'}=\delta_{\sigma\sigma'}\Gamma_{\alpha\sigma}$.
$\Gamma_{\alpha\uparrow}\ge\Gamma_{\alpha\downarrow}\ge 0$ are the positive eigenvalues
which can be written as $\Gamma_{\alpha\sigma}=\Gamma_\alpha \frac{1}{2} (1+\sigma p_\alpha)$,
with $\Gamma_\alpha\ge 0$ and $0\le p_\alpha \le 1$. Defining
$\Gamma_{\alpha\sigma}=2\pi t_{\alpha\sigma}^2$ and 
$\Gamma_\alpha=4\pi t_\alpha^2$, with $t_\alpha,t_{\alpha\sigma}\ge 0$, 
we can write $\dunderline{\Gamma}{}_\alpha$ in the two equivalent forms
\begin{align}
\label{eq:hybridization_matrix_form_1}
\dunderline{\Gamma}{}_{\alpha}&=2\pi t_\alpha^2
\dunderline{\rho}{}_\alpha \quad,\quad
\dunderline{\rho}{}_\alpha=
\dunderline{U}{}_\alpha^\dagger(2\dunderline{\tilde{\Gamma}}{}_\alpha/\Gamma_\alpha)
\dunderline{U}{}_\alpha\,,\\
\label{eq:hybridization_matrix_form_2}
\dunderline{\Gamma}{}_{\alpha}&=2\pi \dunderline{t}{}_\alpha^\dagger 
\dunderline{t}{}_\alpha \quad,\quad
(\dunderline{t}{}_\alpha)_{\sigma\sigma'}=t_{\alpha\sigma} 
(\dunderline{U}{}_\alpha)_{\sigma\sigma'} \,.
\end{align}
The first form is the one where the information is fully shifted to an effective
d.o.s. $\dunderline{\rho}{}_\alpha$ of the reservoirs with spin-conserving tunneling rates $\Gamma_\alpha$.
Using $2\dunderline{\tilde{\Gamma}}{}_\alpha/\Gamma_\alpha=
\dunderline{\mathbbm{1}}+p_\alpha \dunderline{\sigma}^z$
and $\dunderline{U}{}_\alpha=e^{i\frac{1}{2}\vec{\varphi}_\alpha\underline{\underline{\vec{\sigma}}}}$ 
we find $\dunderline{\rho}{}_\alpha=
\dunderline{\mathbbm{1}}+p_\alpha{\hat{\vec{d}}_\alpha}\dunderline{\vec{\sigma}}$,
where $\dunderline{\vec{\sigma}}$ are the Pauli matrices and 
$\hat{\vec{d}}_\alpha=R(\vec{\varphi}_\alpha)\vec{e}_z$ is a unit vector obtained by rotating the $z$-axis 
with rotation axis $\vec{\varphi}_\alpha$. As a result we find an effective model with ferromagnetic
leads with pseudo-spin channels $\sigma=\uparrow,\downarrow$, spin orientation
$\hat{\vec{d}}_\alpha$ and spin polarization $p_\alpha$, see Fig.~\ref{fig:model}. 
Alternatively, one can also shift the whole information into an effective tunneling matrix 
$\dunderline{t}{}_\alpha$, as written in Eq.~\eqref{eq:hybridization_matrix_form_2}, which
describes a model with an effective tunneling matrix and reservoirs without spin polarization.
This will be the form we will use in the following. 

{\it Coulomb blockade regime.} We now present a weak coupling RG analysis close to the particle-hole
symmetric point in the Coulomb blockade regime, defined by 
$D=\epsilon+U=-\epsilon\gg \Lambda_c=\text{max}\{\{|\mu_\alpha|\},h\}$. Charge fluctuations
are suppressed in this regime and, using a Schrieffer-Wolff transformation \cite{korb_etal_prb07},
spin fluctuations are described by the effective interaction 
$V_{\text{eff}}=\sum_{kk'}\uline{a}^\dagger_k \dunderline{\vec{J}}\,\uline{a}_{k'}\vec{S}$,
where $\vec{S}$ denotes the dot spin and 
$\dunderline{\vec{J}}=2\dunderline{t}\,\dunderline{\vec{\sigma}}\,\dunderline{t}^\dagger/D$ 
is an effective exchange matrix. $(\uline{a}_k)_{\alpha\sigma}=a_{k\alpha\sigma}$
is a vector containing all reservoir field operators and 
$(\dunderline{t})_{\alpha\sigma,\sigma'}=(\dunderline{t}{}_\alpha)_{\sigma\sigma'}$
is a matrix containing all tunneling matrices. Via a standard poor man scaling RG analysis we integrate out
all energy scales between $D$ and $\Lambda_c$. In this regime the chemical potentials $\mu_\alpha$ 
do not enter and it is 
convenient to rotate all reservoirs such that only one reservoir
couples effectively to the dot. This is achieved by the singular value decomposition 
$\dunderline{t}=\dunderline{V}\,\dunderline{\tilde{t}}\,\dunderline{W}^\dagger$,
where $\dunderline{V}$ and $\dunderline{W}$ are unitary transformations in reservoir and dot space,
respectively, and 
$(\dunderline{\tilde{t}})_{\alpha\sigma,\sigma'}=\delta_{\alpha 1}\delta_{\sigma\sigma'}
\lambda_\sigma$ contains the two singular values 
$\lambda_\uparrow\ge\lambda_\downarrow > 0$ of the tunneling matrix. We exclude here the 
exotic case $\lambda_\downarrow=0$ which would mean that one of the dot levels 
effectively decouples from the reservoirs. By rotating dot space, we can omit the 
matrix $\dunderline{W}$ and the tunneling matrices are given by
$\dunderline{t}{}_\alpha=\dunderline{V}{}_\alpha\dunderline{\lambda}$ with
$(\dunderline{V}{}_\alpha)_{\sigma\sigma'}=(\dunderline{V})_{\alpha\sigma,\sigma'}$ and 
$(\dunderline{\lambda})_{\sigma\sigma'}=\delta_{\sigma\sigma'}\lambda_\sigma$. 
For the RG we omit the unitary transformation $\dunderline{V}$ such that only 
one effective reservoir couples to the dot via the tunneling matrix elements
$\lambda_\sigma$. This model has also been studied in Ref.~\onlinecite{kashcheyevs_etal_prb07}
and leads to an effective $2\times 2$ exchange coupling matrix 
$\dunderline{\vec{\tilde{J}}}=2\dunderline{\lambda}\,\dunderline{\vec{\sigma}}\,\dunderline{\lambda}/D$
which can be parametrized by two exchange couplings
$J_z=(\lambda_\uparrow^2 + \lambda_\downarrow^2)/D$ and
$J_\perp=2\lambda_\uparrow\lambda_\downarrow/D$ via
$\dunderline{\tilde{J}}^z=c\dunderline{\mathbbm{1}}+J_z\dunderline{\sigma}^z$ and
$\dunderline{\tilde{J}}^{x,y}=J_\perp\dunderline{\sigma}^{x,y}$, with
$c=\sqrt{J_z^2 - J_\perp^2}$ and $J_z\ge J_\perp > 0$. As a result one obtains the
antiferromagnetic anisotropic Kondo model together with a potential scattering
term from the anisotropy constant $c$. The weak-coupling RG flow as function of the
effective band width $\Lambda$ leads to an increase of the exchange couplings
towards the isotropic fixed point $J_z=J_\perp$ with $c$ and $T_K=\Lambda[(J_z-c)/(J_z+c)]^{1/(4c)}$ being
the invariants. At each stage of the RG flow we can replace
$D\rightarrow\Lambda$ and get the effective hybridization matrix 
$\dunderline{\Gamma}{}_\alpha=2\pi\dunderline{\lambda}\,
\dunderline{V}{}_\alpha^\dagger\dunderline{V}{}_\alpha\dunderline{\lambda}$, where $\dunderline{\lambda}$ contains
the renormalized exchange couplings $J_{z,\perp}$ via
$\lambda_{\uparrow,\downarrow}^2=\Lambda(J_z\pm c)/2$. The matrices $\dunderline{V}{}_\alpha$ do not
flow under the RG and fulfill 
$\sum_\alpha \dunderline{V}{}_\alpha^\dagger \dunderline{V}{}_\alpha = \dunderline{\mathbbm{1}}$ since 
$\dunderline{V}$ is unitary. This leads to $\sum_\alpha\dunderline{\Gamma}{}_\alpha=2\pi\dunderline{\lambda}^2$. 
Comparing this to the form 
$\sum_\alpha\dunderline{\Gamma}{}_\alpha=\frac{\Gamma}{2}(\dunderline{\mathbbm{1}}+\vec{d}\dunderline{\vec{\sigma}})$ from
\eqref{eq:hybridization_matrix_form_1} we find $J_z=\Gamma/(2\pi\Lambda)$ and
$d=|\vec{d}|=c/J_z$. We conclude that the system shows a 
tendency to minimize the vector $\vec{d}$ during the RG flow and,
for $c\ll J_z$, we can set this vector to zero and obtain the central result \eqref{eq:univ_prop}.
This is reached in the scaling limit, formally defined in terms of the initial parameters by
$J^{(0)}_{z,\perp}\rightarrow 0$ and $D\rightarrow\infty$ such that the Kondo temperature $T_K$ 
and the ratio $J^{(0)}_z/J^{(0)}_\perp$ are kept fixed. At this isotropic fixed point, we get
$\lambda_1=\lambda_2=\lambda$ and 
$\dunderline{\Gamma}{}_\alpha=2\pi\lambda^2\dunderline{V}{}_\alpha^\dagger\dunderline{V}{}_\alpha$. 
Using the form \eqref{eq:hybridization_matrix_form_1} we find 
$\dunderline{V}{}_\alpha^\dagger\dunderline{V}{}_\alpha=
x_\alpha \dunderline{\mathbbm{1}}+\vec{d}_\alpha\dunderline{\vec{\sigma}}$ providing 
a recipe to find the parameters $x_\alpha$ and $\vec{d}_\alpha$ at the fixed point.

As already explained in the introduction, for reservoirs with different chemical potentials 
$\mu_\alpha$, the fixed point model gives rise to new interesting universal behavior compared to the
Kondo model with unpolarized leads $\vec{d}_\alpha=0$. The latter case is only the
fixed point model when the initial spin vectors are {\it all} equal $\vec{d}^{(0)}_\alpha=\vec{d}^{(0)}$. 
Whereas a small deviation between the initial polarizations $p_\alpha$ will still end
up in a fixed point with $p_{\alpha}\ll 1$, a small angle between the spin orientations 
leads to a rotation of the spin orientations but the polarizations remain finite.
A special case are reservoirs with full spin polarization $p_\alpha^{(0)}=1$ which 
remain fully spinpolarized during the whole RG flow. In conclusion we find that
the Kondo model with unpolarized leads will almost never describe the correct 
universal behavior in nonequilibrium. 

The characteristic features at and away from the fixed point can best be visualized by
analyzing the stationary dot magnetization $\vec{M}$ and the charge current $I$ in the strong 
nonequilibrium regime
$\Lambda_c=\text{max}\{|\mu_\alpha|\}\gg T_K$ as function of the magnetic field $h<\Lambda_c$.
For $h\gg \gamma \sim J_{z,\perp}^2 \Lambda_c$ ($\gamma$ sets the scale of the rates) a
standard golden rule theory is sufficient to calculate $\vec{M}$ and $I/\gamma$ up to
$O(1)$. In this regime $\vec{M}$ is either parallel
or antiparallel to $\vec{h}$ (depending on the nonequilibrium occupations) and the
magnetization perpendicular to the field is negligible of $O(J_{z,\perp}^2)$. 
For $h\lesssim \gamma$ quantum interference
phenomena are very important and golden rule theory breaks down. A strong component of the 
magnetization perpendicular to the magnetic field of $O(1)$ is obtained and the nondiagonal matrix 
elements of the dot density matrix (accounting for a spin component perpendicular to the magnetic 
field) have to be taken into account.
In the supplemental material we present the analytical results for all regimes which can be obtained
from a systematic analysis of the effective dot Liouville operator 
up to $O(J_{z,\perp}^2)$. The full formulas are very involved but can be simplified in certain regimes. Here
we summarize the most important nonequilibrium features.

\begin{figure}[t]
  \centering
  \includegraphics[width=\linewidth]{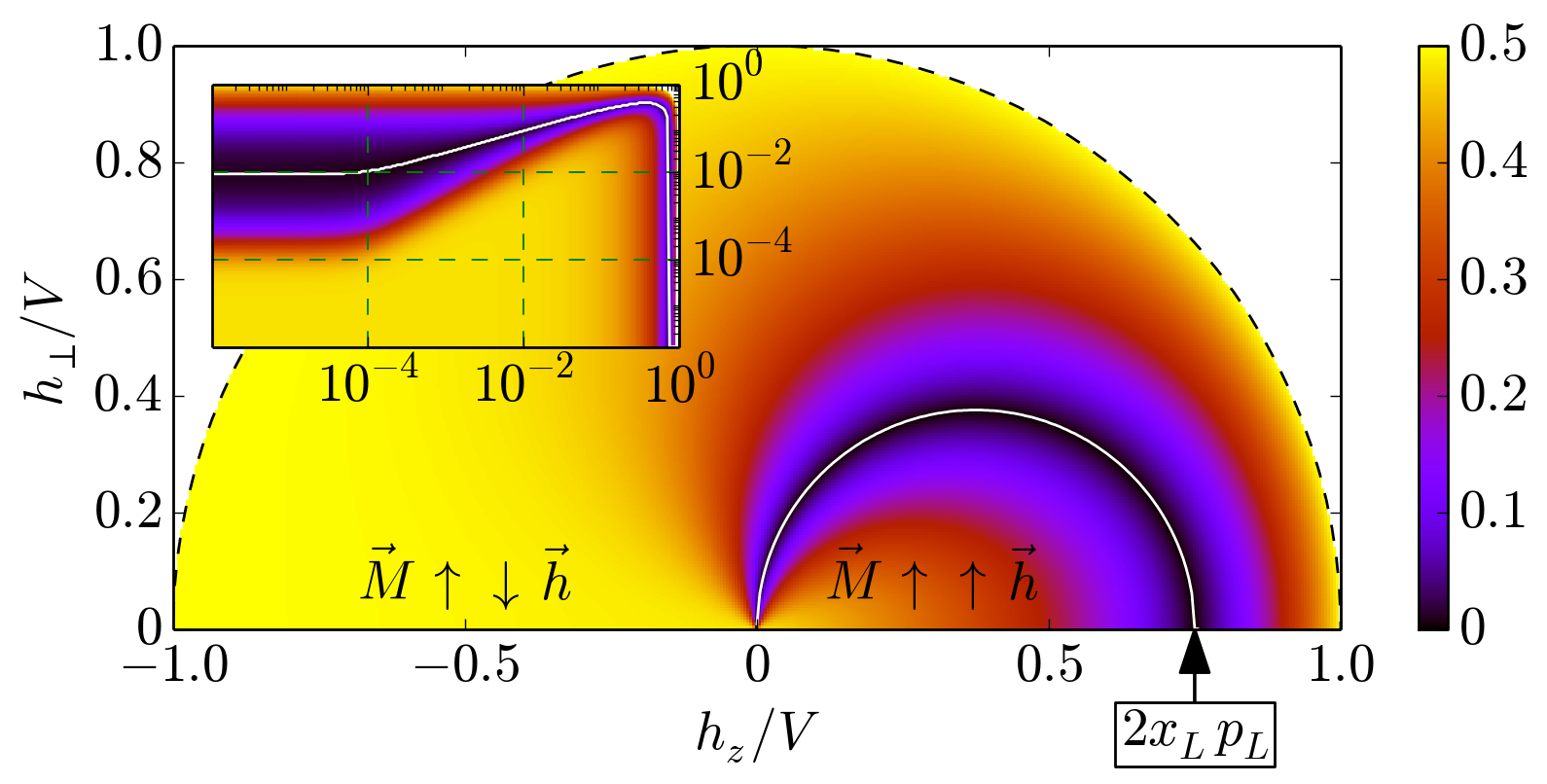}
  \caption{(Color online)
  The dot magnetization $M$ as function of $h_z$ and $h_\perp$ for $h<V$ with $x_L=x_R=\frac{1}{2}$, 
  $p_L=p_R=\frac{3}{4}$, $J =\frac{1}{100\sqrt{\pi}} $ and $\gamma=10^{-4}V$. 
  The white line indicates $h^{\text{min}}_\perp(h_z)$
  where $M$ is minimal. Inset: The same plot on logarithmic scale for $h_z>0$.}
  \label{fig:magnetization}
\end{figure}

\begin{figure}[t]
  \centering
  \includegraphics[width=\linewidth]{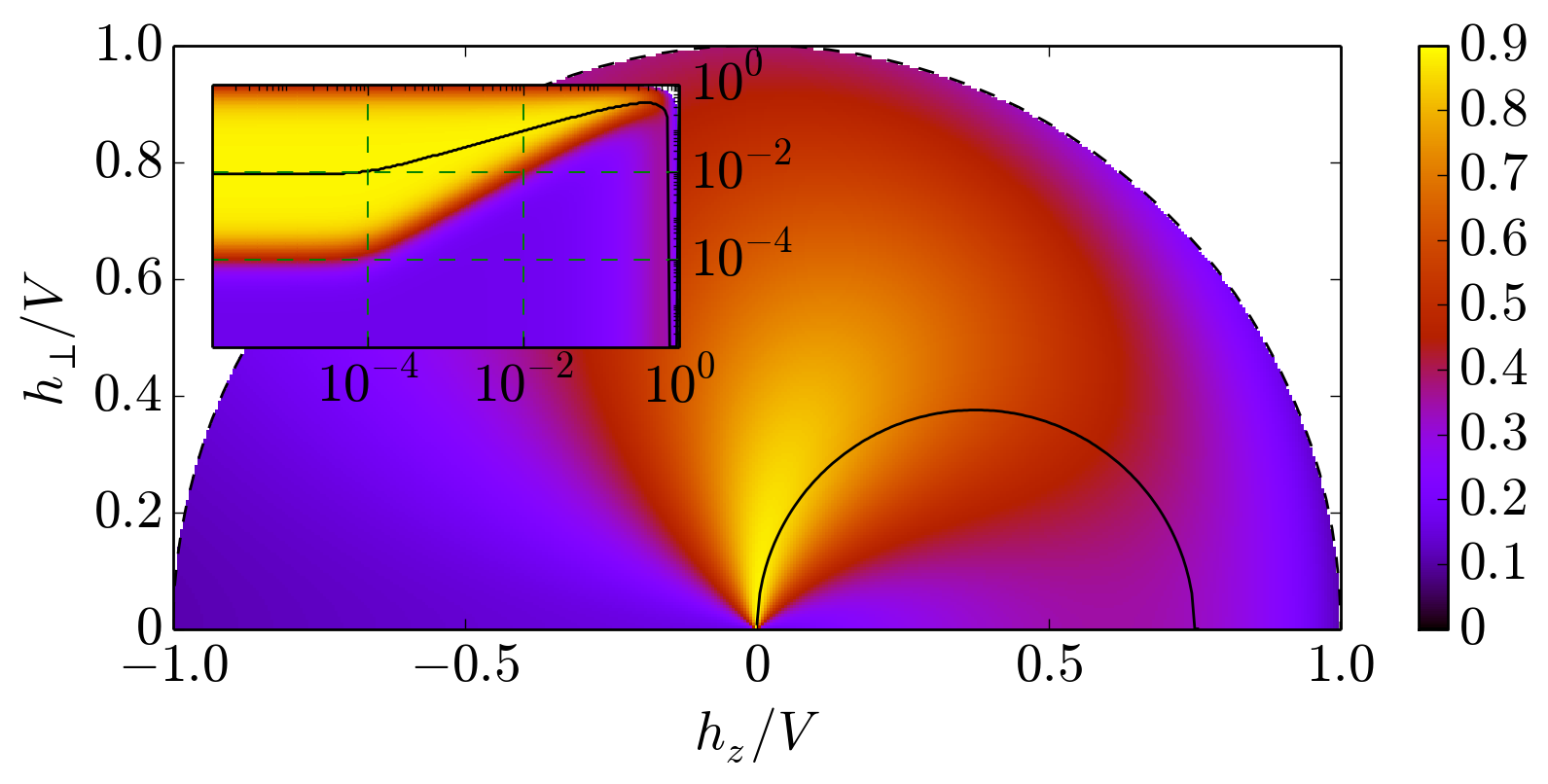}
  \caption{(Color online)
  The charge current $I/\gamma$ in units of the Korringa rate, analog to Fig.~\ref{fig:magnetization}.
  The black line corresponds to the white one of Fig.~\ref{fig:magnetization} indicating minimal $M$.}
  \label{fig:current}
\end{figure}

{\it Dot magnetization in golden rule, arbitrary number of reservoirs at or away from the fixed point.} 
We first start with the regime $h\gg\gamma$ for an arbitrary number of reservoirs. 
The magnetization $\vec{M}$ in
golden rule is zero if the rates between the two spin states are equal. This occurs for magnetic
fields lying on the surface of an ellipsoid which can be fully characterized by the two vectors
$\vec{d}$ and $\vec{\mu}$ defined in Eqs.~\eqref{eq:univ_prop} and \eqref{eq:smoking_gun}, together
with the factor $s=J_z/J_\perp = 1/(1-d^2) \ge 1$ characterizing the distance to the isotropic 
fixed point $s=1$. We find an ellipsoid which is rotationally invariant around $\vec{d}$ and 
stretched along $\vec{d}$ by the factor $s$
\begin{align}
\label{eq:ellipsoid}
(\vec{h}_\perp - \vec{\mu}_\perp)^2 + \left(\frac{h_\parallel - s^2 \mu_\parallel}{s}\right)^2 =
\vec{\mu}_\perp^2 + s^2\mu_\parallel^2 \quad,
\end{align}
where we have decomposed the two vectors $\vec{h}$ and $\vec{\mu}$ in two components parallel and
perpendicular to $\vec{d}$. This result provides an experimental tool to measure the distance to the
fixed point model via the stretching factor $s$ and sets a {\it smoking gun} for a characteristic
universal feature of the fixed point $s=1$, where the ellipsoid turns into the sphere 
\eqref{eq:smoking_gun}. These features are essentially different from the Kondo model with
unpolarized leads where $\vec{d}=\vec{\mu}=0$ such that minimal magnetization in golden rule
occurs only for $\vec{h}=0$. 
We note that at the fixed point the center of the sphere is given by the 
vector $\vec{\mu}$, which is a characteristic vector determining the exchange field 
generated by the reservoirs given by 
$\vec{h}_{\text{exc}}=J(2\vec{\mu}-\vec{h}_{\text{ext}})$, where $\vec{h}_{\text{ext}}$ is the 
externally applied field (this can be obtained by a perturbative calculation similiar to 
the one of Ref.~\onlinecite{koenig_etal}). Outside (inside) the ellipsoid the magnetization is 
antiparallel (parallel)
to $\vec{h}$ but the rotational symmetry around the vector $\vec{d}$ is no longer valid
since all scalar products $\vec{d}_\alpha\vec{h}$ enter. Only in
the special case of two reservoirs $\alpha=L,R$ at the fixed point $\vec{d}_L = -\vec{d}_R$ we 
obtain antiparallel spin orientations of the two reservoirs with rotational symmetry around the 
reservoir spin axis. The universal properties of this case are shown in Fig.~\ref{fig:magnetization} 
for the dot magnetization and in Fig.~\ref{fig:current} for the charge current
and will be discussed in more detail in the following including the quantum interference 
regime $h\lesssim\gamma$.

{\it Dot magnetization, 2 reservoirs at the fixed point.} 
For two reservoirs at the fixed point, we choose $\vec{d}_L=-\vec{d}_R$ in z-direction and 
characterize the coupling $J$ by the Korringa rate $\gamma=4x_L x_R \pi J^2 V$, where $V=\mu_L-\mu_R$
is the bias voltage. From $\vec{\mu}=V\vec{d}_L$ and $|\vec{\mu}|=V x_L p_L$ the minimum of the
magnetization in the golden rule regime $h\gg\gamma$ lies on a sphere centered around 
$h_z=x_L p_L V$, $h_\perp=0$ with radius $x_L p_L V$. Since $2 x_L p_L =2 x_L x_R(p_L+p_R)\le (p_L+p_R)/2\le 1$, 
the sphere will always lie inside the region $h<V$. At $h=V$ we get $\vec{M}=-\vec{h}/(2V)$. These 
features follow from energy conservation and the fact that the majority spins in the left/right lead are
$\uparrow / \downarrow$. For small $h_\perp$ the upper level of the dot consists mainly of the 
spin-$\uparrow$ state which will be occupied from the left lead but has a small probability to escape
to the right one. Therefore the magnetization is parallel to the external field and quite large (but 
not maximal). Increasing $h_\perp$ will lead to transition rates between the upper and
lower dot level until they are equal, which defines the minimum of the magnetization. 
For large $h_\perp\sim O(V)$ the energy phase space for the transition from the lower to the 
upper level becomes smaller
leading to an increase of the population of the lower level. Thus, the magnetization becomes antiparallel
to the magnetic field and the magnitude increases until $h=V$, where only the lower level is occupied 
and the magnetization becomes maximal. For $h_z<0$ this mechanism does not occur
since in this case the lower level will always have a higher occupation. For small magnetic fields
$h\lesssim\gamma$ quantum interference processes become important and the minimum position
of the magnetization saturates at $h^{\text{min}}_\perp(h_z)\sim O(JV)$, see the inset of
Fig.~\ref{fig:magnetization}. For $h_z\lesssim\gamma$ and $h_\perp\ll V$, 
the precise line shape follows from $M\approx\sqrt{\pi^2 J^4 x^2 + M_z^2 (1+x^2)}$ with
\begin{align}
\label{eq:Mz_small_hz}
M_z &\approx \frac{1}{2} \frac{p_L+p_R-2\pi J^2 x^2 h_z/\gamma}{1+p_L p_R + x^2}\,\,,\,\,
x=\frac{h_\perp}{\sqrt{h_z^2+\gamma^2}}\,.
\end{align}
At $h=0$ we obtain $M_0=M_{h=0}=(1/2)(p_L+p_R)/(1+p_L p_R)$ which, together with 
$x_L+x_R=1$, $x_L p_L = x_R p_R$ and the value $x_L p_L$ from the minimum magnetization, 
determines the four parameters $x_{L,R}$ and 
$p_{L,R}$ of the fixed point model. The coupling $J$ is related to the 
Korringa rate which follows from the curvature of the magnetization as function of $h_\perp$ at
the origin: $(\partial^2 M/\partial h_\perp^2)_{h=0}= -\gamma^{-2}M_0(1+p_L p_R)/(1-p_L p_R)$.
Furthermore, for vanishing $h_\perp$, the point $h_z=0$ can be characterized by a jump
of the derivative $(\partial M/\partial h_z)|_{h=0}$ with a ratio given by the 
parameters $x_{L,R}$ and $p_{L,R}$, see supplementary material. 

{\it Charge current, 2 reservoirs at the fixed point.}
The charge current $I$ in units of the Korringa rate is shown in Fig.~\ref{fig:current}. 
The current is related to the magnetization in a universal way by the formula
\begin{align}
\nonumber
(I-I_0)/\gamma &= \vec{M}_\perp\vec{h}_\perp/V \\
\label{eq:current_main}
&\hspace{-1cm} + (1+p_L p_R)(M_z-M_0)(h_z/V-2M_0) \quad,
\end{align}
with $I_0/\gamma=I_{h=0}/\gamma=1/2+(1+p_L p_R)(1-8M_0^2)/4$.
At fixed $h_z$ the current shows a 
maximum as function of $h_\perp$ at a value roughly of the same order
where the magnetization is minimal. This is caused by enhanced inelastic processes
increasing the current in this regime. However, since the current varies only slowly in a wide
region around the maximum this is not useful to determine the model parameters. An exception is the
axis $h_z=0$, where the maximum current follows from the formula 
$I^{\text{max}}_{h_z=0}/\gamma=(3+p_L p_R)/4$.
Another point of interest is $h=V$ where the magnetization is maximal $\vec{M}=-\vec{h}/(2V)$ 
(see above). At this point the upper dot level has no
occupation and transport happens via elastic cotunneling processes through the lower one. 
From Eq.~\eqref{eq:current_main} we get $I_{h=V}/\gamma=[1-p_L p_R(2h_z^2/V^2-1)]/4$.
For $h_z=0$, $h_\perp=V$ or $h_z=V$, $h_\perp=0$ this gives $I/\gamma=(1\pm p_L p_R)/4$. 
These two values are related to $I^{\text{max}}_{h_z=0}$ in a universal
way. Together with $I_0$ the parameters $p_{L,R}$ and $\gamma$ can
be determined and $x_{L,R}$ follow from $x_L+x_R=1$ and $x_L p_L = x_R p_R$.
In the quantum interference regime of small magnetic fields the current is shown in the inset 
of Fig.~\ref{fig:current}. Analytically the features follow for $h_z\lesssim\gamma$ and
$h_\perp\ll V$ from $(I-I_0)/\gamma\approx (p_L+p_R)M_0 x^2/(1+p_L p_R + x^2)$.

{\it Conclusions.} We have shown that the Kondo model with unpolarized leads is generically
not the appropriate model to describe the nonequilibrium properties of 
pseudo-spin-$\frac{1}{2}$ quantum dots in the Coulomb blockade regime. 
Noncollinear spin orientations in effective reservoirs give rise to characteristic
features as function of an applied magnetic field in the strong nonequilibrium regime
independent of the microscopic details of the model, even away from the fixed point.
These features are experimentally accessible. For future research it is of high interest
to characterize the universal properties of the model also in the strong coupling regime 
$V\sim T_K$ where more refined techniques have to be used 
\cite{frg_noneq,strong_coupling_RTRG,smirnov_grifoni_03}.

This work was supported by the DFG via FOR 723 and 912. 
We thank V. Meden, M. Pletyukhov, D. Schuricht, and M. Wegewijs for valuable discussions.

\end{document}

%%%%%%%%%%%%%%%%%%%%%%%%%%%%%%%%%%%%%%%%%%%%%%
\title{{\begin{center}Supplemental material for:\\
The generic fixed point model for 
pseudo-spin-$\frac{1}{2}$ quantum dots in nonequilibrium:
\\
Spin-valve systems with compensating spin polarizations\end{center}}}
%%%%%%%%%%%%%%%%%%%%%%%%%%%%%%%%%%%%%%%%%%%%%%
\author{Stefan G\"ottel}
\author{Frank Reininghaus}
\author{Herbert Schoeller}
\affiliation{Institute for Theory of Statistical Physics, RWTH Aachen,
52056 Aachen, Germany}
\affiliation{JARA-Fundamentals of Future Information Technology}

\date{\today}
\pagestyle{plain}

\maketitle
\onecolumngrid
Here we present the technical details of how to calculate the dot magnetization and the 
charge current in the stationary limit by a 
perturbative treatment in the renormalized exchange couplings. We use the Liouville operator method described 
in Ref.~\onlinecite{schoeller_epj09}. We will show the general scheme for the model 
away from the scaling limit for an arbitrary number of reservoirs and analyze the magnetization in detail.
Furthermore, we subsequently will evaluate the formulas for the special case of two reservoirs in the scaling limit.
\\ \\
{\it \bf{General scheme}.}--- 
After renormalization we have obtained the following effective hybridization matrix
\begin{align}
\label{eq:Gamma_general}
\dunderline{\Gamma}{}_\alpha \,=\, 2\pi\,\dunderline{\lambda}\,\dunderline{V}{}_\alpha^\dagger\,
\dunderline{V}{}_\alpha\,\dunderline{\lambda} 
\quad,\quad
\dunderline{\lambda}^2 \,=\,\frac{1}{2}\,\Lambda_c\,(J_z\,\dunderline{\mathbbm{1}}\,+\,c\,\dunderline{\sigma}^z)
\quad,
\end{align}
where $\Lambda_c$ is the low-energy cutoff, 
$c=\sqrt{J_z^2 - J_\perp^2}$ one invariant of the flow equations, and $J_{z,\perp}$ 
are the renormalized exchange couplings from the anisotropic Kondo model. The matrices
$\dunderline{V}{}_\alpha$ follow from the first two columns of the matrix $\dunderline{V}$, which 
was defined by the singular value decomposition of the full tunneling matrix of the original
model. Since $\dunderline{V}$ is unitary we have the important property
$\sum_\alpha \dunderline{V}{}_\alpha^\dagger \dunderline{V}{}_\alpha=\dunderline{\mathbbm{1}}$. 
The hybridization matrix can be decomposed into the unity and Pauli matrices as
\begin{align}
\label{eq:Gamma_decomposition}
\dunderline{\Gamma}{}_\alpha\,=\,\frac{1}{2}\,\Gamma\,\dunderline{x}{}_\alpha
\quad,\quad
\dunderline{x}{}_\alpha \,=\,x_\alpha \dunderline{\mathbbm{1}}\,+\,\vec{d}_\alpha\,\dunderline{\vec{\sigma}}
\quad,\quad
\sum_\alpha\,x_\alpha\,=\,1
\quad,\quad
|\vec{d}_\alpha|\,=\,x_\alpha p_\alpha
\quad,\quad
0\,\le\,x_\alpha,p_\alpha\,\le\,1
\quad,
\end{align}
where $x_\alpha$, $p_\alpha$, and $\hat{\vec{d}}_\alpha$ describe the parameters of
an effective model with ferromagnetic leads. Using 
$\sum_\alpha \dunderline{V}{}_\alpha^\dagger \dunderline{V}{}_\alpha=\dunderline{\mathbbm{1}}$ 
we obtain the properties
\begin{align}
\label{eq:properties_geometric_parameters}
\Gamma\,=\,2\pi\,\Lambda_c\,J_z \quad,\quad
d\,=\,|\vec{d}|\,=\,\frac{c}{J_z} \quad,\quad
\vec{d}\,=\,\sum_\alpha \vec{d}_\alpha \,=\, d\,\vec{e}_z \quad.
\end{align}
This result shows that the $z$-axis has been chosen such that the vector
$\vec{d}$ points in $z$-direction. This choice
has been taken when omitting the unitary dot matrix $W$ in the singular value decomposition
of the full tunneling matrix. Independent of the scale $\Lambda_c$ the matrix $\dunderline{x}{}_\alpha$ 
containing the effective model parameters can be defined from the equation
\begin{align}
\label{eq:x_matrix_def}
\dunderline{x}{}_\alpha\,=\,x_\alpha\,\dunderline{\mathbbm{1}}\,+\,\vec{d}_\alpha\,
\dunderline{\vec{\sigma}}\,=\,
(\dunderline{\mathbbm{1}}\,+\,\frac{c}{J_z}\dunderline{\sigma}^z)^{1/2}\,\dunderline{V}{}_\alpha^\dagger\,
\dunderline{V}{}_\alpha\,(\dunderline{\mathbbm{1}}\,+\,\frac{c}{J_z}\dunderline{\sigma}^z)^{1/2}\quad.
\end{align}
This provides an algorithm how to determine the parameters of the effective model
from the knowledge of $\dunderline{V}{}_\alpha$ and $J_{z,\perp}$. In the scaling limit, where
$c/J_z\rightarrow 0$, the r.h.s. of this equation contains only 
$\dunderline{V}{}_\alpha^\dagger\dunderline{V}{}_\alpha$.

From \eqref{eq:Gamma_decomposition} we can easily define an effective tunneling matrix via
\begin{align}
\label{eq:tunneling_def}
\dunderline{\Gamma}{}_\alpha\,=\,2\pi\,\dunderline{t}{}_\alpha^\dagger\,\dunderline{t}{}_\alpha
\quad,\quad
\dunderline{t}{}_\alpha\,=\,\sqrt{\Gamma/(4\pi)}\,\,\dunderline{x}{}_\alpha^{1/2}
\,=\,\sqrt{\Lambda_c J_z /2}\,\,\dunderline{x}{}_\alpha^{1/2}
\quad.
\end{align}
From this tunneling matrix we can calculate the effective exchange coupling matrix describing
the model in the Coulomb blockade regime 
\begin{align}
\label{eq:exchange_coupling_matrix}
\dunderline{\vec{J}}{}_{\alpha\alpha'}\,=\,\frac{2}{\Lambda_c}\,
\dunderline{t}{}_\alpha\,\dunderline{\vec{\sigma}}\,\dunderline{t}{}_{\alpha'}^\dagger
\,=\,J_z\,\dunderline{x}{}_\alpha^{1/2}\,\dunderline{\vec{\sigma}}\,\,\dunderline{x}{}_{\alpha'}^{1/2}
\quad.
\end{align}
This coupling matrix defines the input for the following perturbation theory to obtain
the dot magnetization and the charge current in the stationary state.

Using the formalism of Ref.~\onlinecite{schoeller_epj09}, the stationary dot density matrix $\rho$ 
and the stationary charge current $I^\gamma$ flowing from lead $\gamma$ into the quantum dot follow from
\begin{align}
\label{eq:rho_I}
L\,\rho\,=\,0 \quad,\quad I^\gamma\,=\,-i\,\text{Tr}\,\Sigma^\gamma\,\rho \quad,
\end{align}
where $\text{Tr}$ denotes the trace over the two dot states, $L$ is the effective Liouvillian and 
$\Sigma^\gamma$ is the current kernel. $L$ and $\Sigma^\gamma$ are superoperators acting on usual
operators, i.e. they are $4\times 4$-matrices in the Liouville basis 
$\uparrow\uparrow,\downarrow\downarrow,\uparrow\downarrow,\downarrow\uparrow$, whereas $\rho$ 
is an ordinary operator or a $4$-component vector in the Liouville basis. From the matrix 
elements of the density matrix the stationary dot magnetization $\vec{M}=\langle\vec{S}\rangle$ follows as
\begin{align}
\label{eq:dm_magn}
M_x\,=\,\text{Re}\rho_{\uparrow\downarrow}\quad,\quad
M_y\,=\,\text{Im}\rho_{\uparrow\downarrow}\quad,\quad
M_z\,=\,\frac{1}{2}(\rho_{\uparrow\uparrow}-\rho_{\downarrow\downarrow})\quad,
\end{align}
with $\rho_{\uparrow\uparrow}+\rho_{\downarrow\downarrow}=1$ and 
$\rho_{\downarrow\uparrow}=\rho_{\uparrow\downarrow}^*$.

In second order perturbation theory in the exchange coupling matrix, the Liouvillian and the current kernel 
follow from
\begin{align}
\label{eq:Liouvillian}
L\,&=\,\hat{G}_{\alpha\sigma,\alpha'\sigma'}\,\left\{-i\frac{\pi}{2}\,
(\mu_\alpha-\mu_{\alpha'}-L_0)\,\tilde{G}_{\alpha'\sigma',\alpha\sigma}\,+\,
H(\mu_\alpha-\mu_{\alpha'}-L_0)\,\hat{G}_{\alpha'\sigma',\alpha\sigma}\right\}\quad,\\
\label{eq:current_kernel}
\Sigma^\gamma\,&=\,\hat{I}^\gamma_{\alpha\sigma,\alpha'\sigma'}\,\left\{-i\frac{\pi}{2}\,
(\mu_\alpha-\mu_{\alpha'}-L_0)\,\tilde{G}_{\alpha'\sigma',\alpha\sigma}\,+\,
H(\mu_\alpha-\mu_{\alpha'}-L_0)\,\hat{G}_{\alpha'\sigma',\alpha\sigma}\right\}\quad,
\end{align}
where we sum implicitly over all reservoir indizes $\alpha,\alpha'$ and all spin
indizes $\sigma,\sigma'$, and we have defined the superoperator vertices
\begin{align}
\label{eq:G_vertices}
\hat{G}_{\alpha\sigma,\alpha'\sigma'}\,&=\,\vec{J}_{\alpha\sigma,\alpha'\sigma'}\,
(\vec{L}_+\,+\,\vec{L}_-) \quad,\quad
\tilde{G}_{\alpha\sigma,\alpha'\sigma'}\,=\,\vec{J}_{\alpha\sigma,\alpha'\sigma'}\,
(\vec{L}_+\,-\,\vec{L}_-) \quad,\\
\label{eq:I_vertex}
\hat{I}^\gamma_{\alpha\sigma,\alpha'\sigma'}\,&=\,c^\gamma_{\alpha\alpha'}\,
\tilde{G}_{\alpha\sigma,\alpha'\sigma'} 
\quad,\quad c^\gamma_{\alpha\alpha'}\,=\,-\frac{1}{2}\,(\delta_{\alpha\gamma}-\delta_{\alpha'\gamma})
\quad.
\end{align}
$\vec{L}_\pm$ are two dot superoperators defined by the following action on ordinary
dot operators $A$
\begin{align}
\label{eq:L_pm}
\vec{L}_+\,A\,=\,\vec{S}\,A \quad,\quad
\vec{L}_-\,A\,=\,- A\,\vec{S} \quad,
\end{align}
and $\vec{S}$ denotes the spin operator of the dot. $L_0$ and $H(x)$ in Eqs.~\eqref{eq:Liouvillian} and
\eqref{eq:current_kernel} are defined by
\begin{align}
\label{eq:L0_H}
L_0\,=\,\vec{h}\,(\vec{L}_+\,+\,\vec{L}_-) \quad,\quad
H(x)\,=\,x\,\ln \frac{-ix}{\Lambda_c}\,=\,-i\frac{\pi}{2}\,|x|\,+\,x\,\ln{|\frac{x}{\Lambda_c}|}
\,\approx\,-i\frac{\pi}{2}\,|x|\quad.
\end{align}
The logarithmic real part of the function $H(x)$ contributes only to a weak renormalization of the
dot levels and changes the final results only in higher orders. Therefore it is omitted in
the following. Inserting the various definitions into \eqref{eq:Liouvillian} and
\eqref{eq:current_kernel} we obtain
\begin{align}
\label{eq:Liouvillian_explicit}
L\,=\,- i \frac{\pi}{2}\left(\text{Tr}_\sigma\,\dunderline{J}{}_{\alpha\alpha'}^i\,\dunderline{J}{}_{\alpha'\alpha}^j\right)\,
(L_+^i + L_-^i)\,
\left\{(\mu_\alpha-\mu_{\alpha'}-L_0)\,(L_+^j - L_-^j)
\,+\,\left|\mu_\alpha-\mu_{\alpha'}-L_0\right|\,(L_+^j + L_-^j)\right\} \quad,\\
\label{eq:current_kernel_explicit}
\Sigma^\gamma\,=\,- i \frac{\pi}{2} c^\gamma_{\alpha\alpha'}\,
\left(\text{Tr}_\sigma\,\dunderline{J}{}_{\alpha\alpha'}^i\,\dunderline{J}{}_{\alpha'\alpha}^j\right)\,
(L_+^i - L_-^i)\,
\left\{(\mu_\alpha-\mu_{\alpha'}-L_0)\,(L_+^j - L_-^j)
\,+\,\left|\mu_\alpha-\mu_{\alpha'}-L_0\right|\,(L_+^j + L_-^j)\right\} \quad,
\end{align}
where $\text{Tr}_\sigma$ denotes the trace over the reservoir spins in the space of the matrices
$\dunderline{\vec{J}}{}_\alpha$. This trace can be calculated explicitly by using the
form \eqref{eq:exchange_coupling_matrix} of the exchange coupling matrix with $\dunderline{x}{}_\alpha$
given by \eqref{eq:x_matrix_def}. After some algebra one obtains for fixed indizes 
$\alpha,\alpha',i,j$
\begin{align}
\label{eq:trace_sigma}
\text{Tr}_\sigma\,\dunderline{J}{}_{\alpha\alpha'}^i\,\dunderline{J}{}_{\alpha'\alpha}^j\,=\,
2 \,J_z^2\,\left\{
\delta_{ij}\,(x_\alpha x_{\alpha'}-\vec{d}_\alpha\vec{d}_{\alpha'})\,+\,
i(\vec{e}_i\wedge\vec{e}_j)\,(\vec{d}_\alpha - \vec{d}_{\alpha'}) \,+\,
(d_\alpha^i d_{\alpha'}^j + d_\alpha^j d_{\alpha'}^i)\right\}
\quad,
\end{align}
where $\vec{e}_i\wedge\vec{e}_j$ denotes the vector product of the unit vectors $\vec{e}_i$ in
$i$-direction. This expression can be simplified when summing over $\alpha$ or $\alpha'$ by
using the properties \eqref{eq:properties_geometric_parameters} but this is not always possible
when inserting \eqref{eq:trace_sigma} in \eqref{eq:Liouvillian_explicit} and 
\eqref{eq:current_kernel_explicit}.

The formulas \eqref{eq:Liouvillian_explicit}, \eqref{eq:current_kernel_explicit} and
\eqref{eq:trace_sigma} together with \eqref{eq:rho_I} conclude the general formalism
how to obtain the dot magnetization and the charge current for the generic case. The
only task which remains is to insert the $4\times 4$-matrices $\vec{L}_\pm$ and work
out the algebra. This can be done either numerically or analytically. Here we will first
present the analytical result for the general case and then discuss 
the special case of two reservoirs in the scaling limit in more detail.
\\ \\
{\it \bf{Arbitrary number of reservoirs away from the scaling limit}.}--- 
Starting from the rotational invariant equations \eqref{eq:Liouvillian_explicit}, 
\eqref{eq:current_kernel_explicit} and \eqref{eq:trace_sigma} we choose the basis such 
that the magnetic field is parallel or antiparallel to the z-direction 
$h_z = \sigma h$, $h_x = h_y = 0$ with $\sigma = \pm$.
In matrix form we represent the Liouvillian as
\begin{align}
\label{eq:Liouville basis}
L\,=\,\begin{array}{cl} \uparrow\uparrow \\ \downarrow\downarrow \\ 
\uparrow\downarrow \\ \downarrow\uparrow \end{array}
\left(\begin{array}{cc|cc}  \hspace{1cm} & & \hspace{1cm} &  \\ & & & \\ \hline & & & \\ & & &   \end{array}\right)
\quad.
\end{align}
After a lengthy algebra, we obtain
\begin{align*}
  L \,=\,& \begin{pmatrix} A & B \\ C & D \end{pmatrix},\\
  A \,=\,& -i \Gamma_1 \left( \mathbbm{1} - \sigma_x \right) + i \Gamma_2 \left( \sigma_z + i \sigma_y \right),\\
  B \,=\,& - \gamma_1 \left( \mathbbm{1} - \sigma_x \right) + i \gamma_2 \left( \sigma_z + i \sigma_y \right),\\
  C \,=\,& \gamma_1 \left( \mathbbm{1} + \sigma_x \right) + i \gamma_2 \left( \sigma_z - i \sigma_y \right) + i \gamma_3 \left(\mathbbm{1} + \sigma_x \right) + \gamma_4 \left( \sigma_z + i \sigma_y \right),\\
  D \,=\,& \sigma h \sigma_z - i \gamma_5 \mathbbm{1} + i \gamma_6 \sigma_x + i \gamma_7 \sigma_y,
\end{align*}
with the parameters
\begin{align*}
  \Gamma_1 / \left( \pi J_z^2 \right) \,=\,& \frac{1}{2} \left( x_\alpha x_{\alpha'} - d_\alpha^z d_{\alpha'}^z \right) \left(|\mu_{\alpha \alpha'} + h| + |\mu_{\alpha \alpha'} - h|\right) - \sigma d_\alpha^z x_{\alpha'} \left( |\mu_{\alpha \alpha'} + h| - |\mu_{\alpha \alpha'} - h| \right) ,\\
  \Gamma_2 / \left( \pi J_z^2 \right) \,=\,& 2 d_\alpha^z x_{\alpha'} \mu_{\alpha \alpha'} - \sigma h \left( 1 - (d^z)^2 \right) ,\\
  \gamma_1 / \left( \pi J_z^2 \right) \,=\,& \frac{\sigma}{2} d_\alpha^y x_{\alpha'} \left( |\mu_{\alpha \alpha'} + h| - |\mu_{\alpha \alpha'} - h| \right) + d_\alpha^y d_{\alpha'}^z \left( |\mu_{\alpha \alpha'} + h| + |\mu_{\alpha \alpha'} - h| \right) ,\\
  \gamma_2 / \left( \pi J_z^2 \right) \,=\,& \frac{\sigma}{2} d_\alpha^x x_{\alpha'} \left( |\mu_{\alpha \alpha'} + h| - |\mu_{\alpha \alpha'} - h| \right) + d_\alpha^x d_{\alpha'}^z \left( |\mu_{\alpha \alpha'} + h| + |\mu_{\alpha \alpha'} - h| \right) ,\\
  \gamma_3 / \left( \pi J_z^2 \right) \,=\,& 2 d_\alpha^x x_{\alpha'} \mu_{\alpha \alpha'} + \sigma d^x d^z h ,\\
  \gamma_4 / \left( \pi J_z^2 \right) \,=\,&  2 d_\alpha^y x_{\alpha'} \mu_{\alpha \alpha'} + \sigma d^y d^z h ,\\
  \gamma_5 / \left( \pi J_z^2 \right) \,=\,& \left( x_{\alpha} x_{\alpha'} - d_\alpha^z d_{\alpha'}^z \right) |\mu_{\alpha \alpha'}| + \frac{1}{2} \left[ x_{\alpha} x_{\alpha'} - \left( d_\alpha^x d_{\alpha'}^x + d_\alpha^y d_{\alpha'}^y - d_\alpha^z d_{\alpha'}^z \right) \right]\left( |\mu_{\alpha \alpha'} + h| + |\mu_{\alpha \alpha'} - h| \right) ,\\
  \gamma_6 / \left( \pi J_z^2 \right) \,=\,& \left( d_\alpha^x d_{\alpha'}^x - d_\alpha^y d_{\alpha'}^y \right) |\mu_{\alpha \alpha'}| ,\\
  \gamma_7 / \left( \pi J_z^2 \right) \,=\,&  2 d_\alpha^x d_{\alpha'}^y |\mu_{\alpha \alpha'}|.
\end{align*}
Considering the eigenvector with eigenvalue zero of the Liouvillian and using the equations \eqref{eq:dm_magn} the magnetization reads
\begin{align*}
  M_z =& \frac{\Gamma_2 \left( h^2 + \gamma_5^2 - \gamma_6^2 - \gamma_7^2 \right) + 2 \left[ - \gamma_2 \gamma_3 \left( \gamma_5 - \gamma_6 \right) + \gamma_2 \gamma_4 \left( \gamma_7 - \sigma h \right) - \gamma_1 \gamma_4 \left( \gamma_5 + \gamma_6 \right) + \gamma_1 \gamma_3 \left( \gamma_7 + \sigma h \right) \right]}{2 \Gamma_1 \left( h^2 + \gamma_5^2 - \gamma_6^2 - \gamma_7^2\right) + 4 \left[ \gamma_2^2 \left( \gamma_5 - \gamma_6 \right) - 2 \gamma_1 \gamma_2 \gamma_7 + \gamma_1^2 \left( \gamma_5 + \gamma_6 \right) \right]}\\
  M_x =& \frac{\gamma_3 \left( \gamma_5 - \gamma_6 \right) - \gamma_4 \left( \gamma_7 - \sigma h \right) + 2 M_z \left[ \gamma_2 \left( \gamma_5 - \gamma_6 \right) - \gamma_1 \left( \gamma_7 - \sigma h \right) \right]}{h^2 + \gamma_5^2 - \gamma_6^2 - \gamma_7^2}\\
  M_y=& \frac{-\gamma_4 \left( \gamma_5 + \gamma_6 \right) + \gamma_3 \left( \gamma_7 + \sigma h \right) + 2 M_z \left[ - \gamma_1 \left( \gamma_5 + \gamma_6 \right) + \gamma_2 \left( \gamma_7 + \sigma h \right) \right]}{h^2 + \gamma_5^2 - \gamma_6^2 - \gamma_7^2}
\end{align*}
and the current
\begin{align*}
  I^\gamma =& -\pi c_{\alpha \alpha'}^\gamma J_z^2 \Big\{ \left( 3 x_{\alpha} x_{\alpha'} - \vec{d}_\alpha \cdot \vec{d}_{\alpha'} \right) \mu_{\alpha \alpha'} - 4 d_\alpha^z x_{\alpha'} \sigma h \\
  & \qquad + 2 \sigma M_z \left( x_{\alpha} x_{\alpha'} - d_\alpha^z d_{\alpha'}^z \right) \left( |\mu_{\alpha \alpha'} + h| - | \mu_{\alpha \alpha'} - h| \right) - 4 M_z d_\alpha^z x_{\alpha'} \left( |\mu_{\alpha \alpha'} + h| + |\mu_{\alpha \alpha'} - h| \right)\\
  & \qquad- 2 \sigma M_x d_\alpha^x d_{\alpha'}^z \left( |\mu_{\alpha \alpha'} + h| - |\mu_{\alpha \alpha'} - h| \right) - 2 M_x d_\alpha^x x_{\alpha'} \left( |\mu_{\alpha \alpha'} + h| + |\mu_{\alpha \alpha'} - h| + |\mu_{\alpha \alpha'}| \right) \\
  & \qquad + 2 \sigma M_y d_\alpha^y d_{\alpha'}^z \left( |\mu_{\alpha \alpha'} + h| - |\mu_{\alpha \alpha'} - h| \right) + 2 M_y d_\alpha^y x_{\alpha'} \left( |\mu_{\alpha \alpha'} + h| + |\mu_{\alpha \alpha'} - h| + |\mu_{\alpha \alpha'}| \right) \Big\}
\end{align*}
In the regime $|h_z| \gg \gamma\sim J^2_{z,\perp}\Lambda_c$ these results agree with the ones of golden rule. Going back to the rotational invariant case, so no specification of the direction of the magnetic field is assumed, the results are
\begin{align*}
  \vec{M} =& \frac{-h \left( 1 - (\vec{d} \cdot \vec{\phi})^2 \right) + 2 \vec{\mu} \cdot \vec{\phi} }{ \left( x_{\alpha}x_{\alpha'} - \vec{d}_\alpha \cdot \vec{\phi} \;\; \vec{d}_{\alpha'} \cdot \vec{\phi} \right) \left( |\mu_{\alpha \alpha'} + h| + |\mu_{\alpha \alpha'} - h| \right) - 2 \vec{d}_\alpha \cdot \vec{\phi} \,x_{\alpha'} \left( |\mu_{\alpha \alpha'} + h| - |\mu_{\alpha \alpha'} - h| \right) }\vec{\phi}
\end{align*}
with $\vec{\phi} = \vec{h} / h$. 
Choosing the basis such that $\vec{d} = d \vec{e}_z$ with $d = |\vec{d}|$ and using the decomposition $\vec{\mu} = \sum_{i=x,y,z} \mu_i \vec{e}_i$ one can easily verify that the root of the magnetization is given by an ellipsoid equation
\begin{align*}
  \left( h_x - \mu_x \right)^2 + \left( h_y - \mu_y \right)^2 + 
  \left( \frac{h_z - s^2 \mu_z}{s} \right)^2
  =& \,\mu_x^2 + \mu_y^2 + s^2 \mu_z^2,
\end{align*}
where $s = 1/\sqrt{1-d^2}$ is the stretching factor of the ellipsoid in the direction of $\vec{d}$,
which with the help of Eq.~\eqref{eq:properties_geometric_parameters} can be written as
\begin{align*}
  s =  \frac{1}{\sqrt{1-d^2}} = \frac{1}{\sqrt{1-c^2/J_z^2}} =\frac{J_z}{J_\perp} \ge 1 .
\end{align*} 
In the scaling limit $d \rightarrow 0$ (or $J_z=J_\perp$) we get $s=1$ and the ellipsoid becomes a sphere
defined by the equation
\begin{align*}
  |\vec{h}-\vec{\mu}| = |\vec{\mu}| .
\end{align*}
This result is the ``smoking gun'' to determine the distance to the scaling limit.
\\ \\
{\it \bf{Two reservoirs in the scaling limit}.}--- 
For two reservoirs in the scaling limit we consider $J_z=J$, $c=0$, 
$\hat{\vec{d}}_L=-\hat{\vec{d}}_R=\vec{e}_z$, and $x_L p_L = x_R p_R$. Choosing the magnetic field as $\vec{h}=(h_x,0,h_z)$ with $h=|\vec{h}|=\sqrt{h_x^2+h_z^2}<V$, 
we obtain the following result for the Liouvillian 
\begin{align}
\label{eq:Liouville matrix}
L\,=\,
\left(\begin{array}{c|c}  -i\,\Gamma'_1\,(\mathbbm{1}-\sigma^x)\,+\,i\,\Gamma'_2\,(\sigma^z+i\sigma^y)
& -\frac{1}{2}\,h_x\,(\mathbbm{1}-\sigma^x)\,+\,i\,\gamma'_2\,(\sigma^z+i\sigma^y)  
\\ \hline 
-\frac{1}{2}\,h_x\,(\mathbbm{1}-\sigma^x)\,+\, 
i\,\gamma'_2\,(\sigma^z-i\sigma^y)\,+\,i\,\gamma'_3\,(\mathbbm{1}+\sigma^x) & 
h_z\,\sigma^z\,-\,i\,\gamma'_5\,\mathbbm{1}\,+\,i\gamma'_6\,\sigma^x   \end{array}\right)
\quad,
\end{align}
with
\begin{align}
\Gamma'_1/(\pi J^2)\,&=\,
2 x_L x_R (1+p_L p_R)V
\,+\,\frac{1}{2}\sum_\alpha x_\alpha^2(1-p_\alpha^2)(h+h_z^2/h)
\,-\,2 x_L x_R (p_L + p_R)h_z
\quad,\\
\Gamma'_2/(\pi J^2)\,&=\,
2 x_L x_R (p_L + p_R)V
\,-\,h_z
\quad,\\
\gamma'_2/(\pi J^2)\,&=\,
x_L x_R (p_L + p_R)h_x
\,-\,\frac{1}{2}\sum_\alpha x_\alpha^2(1-p_\alpha^2)h_z h_x/h
\quad,\\
\gamma'_3/(\pi J^2)\,&=\,-h_x
\quad,\\
\gamma'_5/(\pi J^2)\,&=\,
4 x_L x_R V
\,-\,\frac{1}{2}\sum_\alpha x_\alpha^2(1-3p_\alpha^2)h_z^2/h
\,+\,\frac{1}{2}\sum_\alpha x_\alpha^2(3-p_\alpha^2)h
\quad,\\
\gamma'_6/(\pi J^2)\,&=\,
-\frac{1}{2}\sum_\alpha x_\alpha^2(1+p_\alpha^2)h_x^2/h
\quad.
\end{align}
Determining the eigenvector with eigenvalue zero of the Liouvillian gives the 
following result for the magnetization
\begin{align}
\label{eq:magnetization_x}
M_x\,&=\,\frac{1}{h_z^2 + (\gamma'_5)^2 - (\gamma'_6)^2}\,
\left\{h_z h_x M_z \,+\,2(\gamma'_5+\gamma'_6)\gamma'_2 M_z \,+\,(\gamma'_5+\gamma'_6)\gamma'_3\right\} 
\quad,\\
\label{eq:magnetization_y}
M_y\,&=\,\frac{1}{h_z^2 + (\gamma'_5)^2 - (\gamma'_6)^2}\,
\left\{(\gamma'_5-\gamma'_6) h_x M_z \,-\,2\gamma'_2 h_z M_z \,-\,\gamma'_3 h_z \right\} 
\quad,\\
\label{eq:magnetization_z}
M_z\,&=\,\frac{\Gamma'_2(h_z^2 + (\gamma'_5)^2 - (\gamma'_6)^2)\,+\,\gamma'_3 h_z h_x \,+\,
2(\gamma'_5+\gamma'_6)\gamma'_2\gamma'_3}{
2\Gamma'_1(h_z^2 + (\gamma'_5)^2 - (\gamma'_6)^2)\,-\,4\gamma'_2 h_z h_x \,+\,
(\gamma'_5-\gamma'_6)h_x^2\,-\,4(\gamma'_5+\gamma'_6)(\gamma'_2)^2}
\quad.
\end{align}
For the current we get the result
\begin{align}
\nonumber
I/\gamma\,&=\,\frac{1}{4}(3+p_L p_R)\,-\,(p_L + p_R)M_z \,+\,M_x h_x/V
\,-\,\frac{1}{2}(p_L + p_R)h_z/V\,+\,(1+p_L p_R)M_z h_z/V \\
\label{eq:current}
&=\,I_0/\gamma\,+\,\vec{M}_\perp\vec{h}_\perp / V
\,+\,(1+p_L p_R)\,(M_z-M_0)\,(h_z/V - 2 M_0)
\quad,
\end{align}
where $\vec{M}_\perp$ and $\vec{h}_\perp$ are the components perpendicular to the 
$z$-axis, $\gamma=4 x_L x_R \pi J^2 V$ is the Korringa rate, and
\begin{align}
\label{eq:current_magnetization_h=0}
M_0\,=\,M_{h=0}\,=\,\frac{1}{2}\,\frac{p_L + p_R}{1 + p_L p_R} 
\quad,\quad
I_0/\gamma\,=\,I_{h=0}/\gamma\,=\,
\frac{1}{4}(3+p_L p_R)\,-\,(p_L + p_R)M_0
\quad.
\end{align}
These complicated formulas can be simplified in certain regimes. For $h_z\gg\gamma$ or $h_x/V \sim O(1)$
or $h_x\ll\gamma$ the magnetization is parallel or antiparallel to the magnetic field and agrees with
the golden rule result
\begin{align}
\label{eq:M_golden_rule}
\vec{M}\,\approx\,\vec{h}\,
\frac{1}{2}\,\frac{2 x_L x_R (p_L + p_R) V h_z\,-\,h^2}{
2 x_L x_R V (h^2 + p_L p_R h_z^2)\,+\,h^3\,-\,2 x_L x_R h (h+p_L h_z)(h+p_R h_z)}
\quad.
\end{align}
This shows that the magnetization has a minimum for 
$2 x_L x_R (p_L + p_R) V h_z=2x_L p_L V h_z=h_z^2+(h_x^{\text{min}}(h_z))^2$, which defines a circle of radius
$x_L p_L V$ centered at $h_z = x_L p_LV$, $h_x=0$ for $h_x^{\text{min}}(h_z)$. For the special
case $h_x=0$ we obtain the result $M_x=M_y=0$ and
\begin{align}
\label{eq:hx=0}
M_z\,=\,\frac{1}{2}\,\frac{2 x_L x_R (p_L+p_R)V-h\,\text{sign}(h_z)}{
h + 2 x_L x_R (1+p_L p_R) (V-h) \,-\, 2x_L x_R (p_L + p_R) h \,\text{sign}(h_z)}
\quad.
\end{align}
This gives rise to a characteristic jump of the first derivative of the magnetization at $h_z=0$ with the ratio
\begin{align}
\label{eq:partial_hz_jump}
\frac{\frac{\partial M }{\partial h_z}\big|_{h_z=0^+,h_x=0}}{
\frac{\partial M}{\partial h_z}\big|_{h_z=0^-,h_x=0}}\,=\,
\frac{(1+p_L)(1+p_R)}{(1-p_L)(1-p_R)}\,\cdot\,\frac{1-2 x_L x_R (p_L + p_R)}{1+2 x_L x_R (p_L + p_R)}
\quad.
\end{align}

For small magnetic fields $h_z \lesssim \gamma $ and $J^2 V \lesssim h_x \lesssim JV$ 
golden rule theory breaks down and a finite component $M_y$ appears. We obtain
for $h_z\sim\gamma$ and $h_x\ll V$
\begin{align}
\label{eq:hz=J2}
M_x\approx\frac{1}{2}\,\frac{h_z h_x}{h_z^2 + \gamma^2}\,
\frac{p_L+p_R-\frac{h_x^2}{2 x_L x_R h_z V}}{1+p_L p_R + x^2}
\quad,\quad 
M_y\approx\frac{1}{2}\,\frac{\gamma h_x}{h_z^2 + \gamma^2}\frac{p_L+p_R}{1+p_L p_R + x^2}
\quad,\quad 
M_z\approx\frac{1}{2}\,\frac{p_L + p_R -\frac{h_z x^2}{2 x_L x_R V}}{1 + p_L p_R + x^2}
\quad,
\end{align}
where $x=h_x/\sqrt{h_z^2+\gamma^2}$.
The same formulas can also be used for $h_z\ll\gamma$ and $h_x\ll V$, except for the case
$h_z\ll\gamma$ and $h_x\lesssim\gamma$, where $M_x$ has to be changed to
\begin{align}
\label{eq:hz_very_small}
M_x\,\approx\,\frac{1}{4 x_L x_R}\,\frac{h_x}{V}\,
\left\{-1\,+\,\frac{x_L x_R (p_L + p_R)^2}{1+p_L p_R + x^2}\right\}\quad.
\end{align}
However, since $M_x\lesssim O(J^2)$ in this regime, it can be neglected. Therefore, we get
in all cases for $h_z\lesssim\gamma$ and $h_x\ll V$ for the total magnetization
\begin{align}
\label{eq:total_magnetization}
M\,\approx\,|\vec{M}|\,\approx\,\sqrt{\pi^2 J^4 x^2\,+\,M_z^2\,(1+x^2)}
\quad,\quad 
M_z\approx\frac{1}{2}\,\frac{p_L + p_R -\frac{h_z x^2}{2 x_L x_R V}}{1 + p_L p_R + x^2}
\quad.
\end{align}

The current in the golden rule regime can easily be evaluated by inserting 
\eqref{eq:M_golden_rule} in \eqref{eq:current}. For small magnetic fields $h_z\lesssim\gamma$
and $h_x\ll V$, we can neglect all terms in the current formula \eqref{eq:current} which
are proportional to the magnetic fields yielding 
$(I-I_0)/\gamma\approx -2M_0(1+p_L p_R)(M_z-M_0)=-(p_L+p_R)(M_z-M_0)$.
Inserting $M_z$ from \eqref{eq:hz=J2} one can see that the term $\sim h_z x^2/V$ in the numerator
can be neglected since it plays only an important role for $x^2\sim V/h_z \gtrsim 1/J^2$ where
$M_z\sim O(J^2)$ can anyhow be neglected. Therefore we obtain for all $h_z\lesssim\gamma$ and
$h_x\ll V$ the result
\begin{align}
\label{eq:current_small_h}
(I-I_0)/\gamma\,\approx\,
-\frac{1}{2}\,(p_L+p_R)^2\,\left(\frac{1}{1+p_L p_R + x^2}\,-\,\frac{1}{1+p_L p_R}\right)
\,=\,\frac{(p_L + p_R)\,M_0\,x^2}{1 + p_L p_R + x^2}
\quad.
\end{align}